\documentclass{osa-article}

\journal{osac}
\newcommand{\p}{\partial}
\newcommand{\ep}{\varepsilon}

\newcommand{\om}{\omega}
\newcommand{\nn}{\nonumber}
\newcommand{\ta}{\theta}

\newcommand{\al}{\alpha}
\newcommand{\cN}{{\cal N}}
\newcommand{\cR}{{\cal R}}
\newcommand{\cH}{{\cal H}}
\newcommand{\cF}{{\cal F}}
\newcommand{\cE}{{\cal E}}
\newcommand{\cW}{{\cal W}}
\newcommand{\td}{{\delta}'_\mu}

\newcommand{\wh}{\widehat}
\newcommand{\be}{\begin{equation}}                                       
\newcommand{\ee}{\end{equation}}
\newcommand{\ba}{\begin{eqnarray}}
\newcommand{\ea}{\end{eqnarray}}
\newcommand{\bl}{\begin{align}}
\newcommand{\el}{\end{align}}
\newcommand{\bref}[1]{(\ref{#1})}

\begin{document}

\title{A hierarchy of coupled mode and envelope models 
	for bi-directional microresonators with Kerr nonlinearity}

\author{Dmitry V. Skryabin\authormark{*}}

\address{Department of Physics, University of Bath, Bath BA2 7AY, UK\\
Russian Quantum Centre, Skolkovo 121205, Russia }

\email{\authormark{*}d.v.skryabin@bath.ac.uk} %% email address is required



\begin{abstract}
We consider interaction of counter-propagating waves in a bi-directionally pumped ring microresonator with Kerr nonlinearity. We introduce a hierarchy of the mode expansions and envelope functions evolving on different 
time scales set by the cavity linewidth and nonlinearity,  dispersion, and repetition rate, and provide a detailed derivation of the corresponding hierarchy of the coupled mode and of the Lugiato-Lefever-like equations. An effect of the washout of the repetition rate frequencies from the equations governing dynamics of the counter-propagating waves is elaborated in details.
\end{abstract}

%%%%%%%%%%%%%%%%%%%%%%%%%%  body  %%%%%%%%%%%%%%%%%%%%%%%%%%

\section{Introduction}
Microresonator frequency combs have been attracting a significant recent attention with their numerous 
practical applications  and as an experimental setting to study 
fundamental physics of dissipative optical solitons, see \cite{rev2,pasquazi,rev1} for recent reviews. 
So called Lugiato-Lefever (LL) model  has become a paradigm in this research area \cite{rev1,ll,herr,chembo}. 
Its  soliton solutions  have some  pre- and post-  Lugiato-Lefever history in and 
outside the optics context, see, e.g.,
\cite{kerr00,kerr0,kerr1,kerr11,kerr2,kerr3,kerr4,kerr5}. However, the area has exploded after a 
breakthrough experimental demonstration of Ref. \cite{herr}.
In terms of the first principle approach to the Kerr microresonator model development, the decade old work \cite{chembo} has remained a main reference. However, together with experimental progress in the area of Kerr microresonators, the underpinning theory deserves a refreshed outlook. One of the recent challenges has emerged after a series of experiments with birectionally pumped microresonators, where combs and solitons have been observed in counter-propagating waves
\cite{vahala,japan2,gaeta,tobias,lobanov}. 
Bi-directionally pumped and, related dual-ring, microresonators have also been recently studied 
for symmetry breaking \cite{japan1,chirality,pascal1,pascal2,pascal3} and gyroscope \cite{vahala2,vahala3,matsko1,matsko2,matsko3,winful} related effects, including idealised PT-symmetric cases \cite{pt1,pt2,pt3}.  

A variety of models has been reported in the context of 
experiments dealing with a single mode 
operation in each direction \cite{japan1,chirality,pascal1,pascal2,pascal3,dudley}. 
We note, here that studies into single mode bidirectional lasers, laser gyroscopes and symmetry breaking in them 
have history going back to 1980's, see, e.g., \cite{mandel,woerdman,skryabin}.
To interpret  recent soliton experiments, \cite{vahala,tobias} have used models without nonlinear cross-coupling, while \cite{japan2}
has accounted for it. As we will see below, neglecting by the nonlinear cross-coupling was probably a better approach to analyse the experimental measurements under the circumstances, when modelling in neither of \cite{vahala,japan2,tobias} included  the effect of opposing group velocities, i.e., opposite signs of the resonator repetition rates for counter-propagating waves. 

Due to complexity of the problem and diversity of equations both met in literature and the ones that are encountered during first principle analysis of the problem, it appears beneficial to have a detailed reference derivation that can be followed and tailored by a reader. Such 
mathematically transparent and physically motivated derivation that can be readily mapped onto a variety of experimental setups is present below.   Focus of our work is to  identify a  
hierarchy of the mode expansions and envelope functions evolving on different 
time scales set by the cavity linewidth, nonlinearity and 2nd order dispersion 
(slow time scales), and by the repetition rate (fast time scale), which can be used to  derive a hierarchy of the coupled mode and envelope equations. We pay  particular attention to comprehensive explanations of our derivation steps and interpretation of the results. 

\section{Hierarchies of mode expansions and envelope functions}
This Section  introduces physical system and discusses a hierarchy of mode amplitudes and envelope functions accounting for different time-scales. It also outlines plan of work for the rest of the paper.

Maxwell equations written for the electric field components $\cE_\alpha$ using Einstein's notations read as
\be
c^2\p_\alpha\p_{\alpha_1} \cE_{\alpha_1}-c^2\p_{\alpha_1}\p_{\alpha_1} \cE_\alpha
+\p_t^2\int_{-\infty}^{\infty}\ep(t-t',r,\ta,z) \cE_\alpha(t',r,\ta,z)dt'= -\p_t^2\cN_\alpha.\label{max}
\ee
Here $\wh\ep$ is the dielectric response function varying in space and time. It  is assumed to be scalar for the sake of brevity. $\ta\in[0,2\pi)$ is the azimuthal coordinate varying along the ring circumference. 
$z$ axis is perpendicular to the ring, while $r=\sqrt{x^2+y^2}$ measures distance from the ring centre.  $\cN_\alpha$ is the nonlinear part of the material polarization and $c$ is the vacuum speed of light.
We assume 3rd order nonlinearity, so that
\be \cN_\alpha=\chi^{(3)}_{\al\al_1\al_2\al_3}
\cE_{\al_1} \cE_{\al_2}\cE_{\al_3},\label{chi3}\ee
where $\alpha_{1,2,3}$ and $\alpha$ represent either of the three Cartesian projections, $x,y$ or $z$, of a physical quantity they are used with.  An implicit summation is assumed over any repeated $\alpha'$s. 
$\chi^{(3)}_{\alpha\alpha_1\alpha_2\alpha_3}$ is a 4th rank tensor of the third order nonlinear susceptibility, which is taken to be nondispersive (Kleinman condition, i.e., interchangeability of all four indices). 

Electric field vector $\cE_\al$ inside a ring resonator is expressed as a superposition of its linear  modes $F_{\alpha j}(r,z)e^{ij\ta\pm i\om_jt}$, which are solutions of  Eq. \bref{max} with $\cN_\alpha=0$:
\be\label{field}
\cE_\alpha =
\sum_{j=j_{min}}^{j_{max}}b_jF_{\alpha j}e^{ij\ta}C_j(t)+c.c., ~
C_j\equiv B^+_j(t)e^{-i\om_jt}+
B^-_j(t)e^{i\om_jt}.
\ee
Here 
$j>0$ is an azimuthal mode number, or angular momentum, and $\om_j>0$ is the corresponding mode frequency. $B^{\pm}_j$ are the  amplitudes of the clockwise (CW) and  counter-clockwise (CCW) modes. For typical microresonators geometries, either bulk crystalline or chip integrated, the transverse mode profiles $F_{\alpha j}$ can be divided into quasi-TE quasi-radial modes ($|F_{x j,yj}|\gg |F_{z j}|$) and quasi-TM ($|F_{z j}|\gg |F_{r j,\ta j}|$). For many practical purposes, which is in our case calculation of the overlap integrals in the nonlinear terms,  it often suffices to neglect the smaller components of $F_{\al j}$. We also assume that the dominant components of $F_{\al j}$ and $\om_j$ are real, so that  $F_{\alpha j}=F_{\alpha j}^*$, $\om_j=\om_j^*$. Thus for TE modes $F_{xj}\approx \cos\ta F_{r}$, $F_{yj}\approx \sin\ta F_{r}$, $F_{zj}\approx 0$ and for TM modes   $F_{xj,yj}\approx 0$.  

In order to cut notational complexity and drop the $\al$ index, we consider TM family, so that from now on $F_{z j}\to F_j$, and
\begin{equation}
\cN_z=\cN=\chi^{(3)}\cE_z^3.
\label{n3}
\end{equation}  
 Results of our derivations would be the same for  TE modes, $F_{r j}\to F_j$, and therefore, what we are loosing is only formal consideration of the nonlinear coupling between the TE and TM families. 
 
We assume that inhomogeneities of the resonator surfaces result in scattering in general and in  backscattering, in particular, and hence lead to the linear coupling between the modes. We account for these effects assuming 
\begin{equation}
\ep(t,\ta, r,z)=\ep_{id}(t,r,z)\big(1+\ep_{in}(\ta)\big).
\end{equation}
 Here $\ep_{id}$
is the dispersive dielectric function of the {\em ideal} (no backscattering) geometry, that does not depend on $\ta$, while  relatively small $\ep_{in}(\ta)$ accounts for {\em inhomogeneities} along the ring. 
Mode profiles $F_{j}(r,z)$ are calculated for  $\ep_{in}=0$. 

$\cE$ is measured in  V/m, hence  normalising linear modes as $\max_{r,z} |F_{ j}|=1$ makes units of  $b_jB^{\pm}_j$ to be V/m. 
Real field amplitude of a CCW mode is $2b_{j}|B^+_{j}|$, so that its intensity is
$I^+_j=2c\epsilon_{vac}n_{j}b_{j}^2|B^+_{j}|^2$ and power is $I^+_j/S_j$.  $S_j$ is the 
effective transverse mode area, $S_j=\big(\iint |F_j^\prime|^2 dxdz\big)^2/\iint |F_j^\prime|^4 dxdz$ and $F_j^\prime=F_j(r,z)\vert_{y=0}$.  $n_j$ is the linear refractive index, $n_j^2=\int_{-\infty}^\infty\ep_{id}(\tau,r=r_0,z=0)e^{i\om_j\tau}d\tau$, where  $r_0$ is the distance between the $z$ axis and a point of maximum of $|F_j|$.
 We define  scaling factors $b_j$  as
\be 
b_j^2=\frac{1}{2\epsilon_{vac}S_jn_jc},
\label{b}
\ee
so that  the $|B^{\pm}_j|^2$  are measured in Watts.  $\epsilon_{vac}$ is the vacuum susceptibility. 

We assume that the resonator is pumped into its $j_p$ mode and  introduce the mode index offset $\mu=j-j_p$.  The real field expression, Eq. \bref{field}, is then  
\ba
&&
\nn  \cE_z(t,\ta,r,z) \simeq 
b_{j_p}F_{j_p}\cE,\\
&&
\cE(t,\ta)=\sum_\mu\Big(B^+_\mu e^{ij_\mu\ta-i\om_\mu t}+
B^-_\mu e^{ij_\mu\ta+i\om_\mu t}\Big)+c.c.,
\label{field1}
\ea
where $j_\mu=j_p+\mu=j$ and $\mu=-|j_p-j_{min}|,\dots,0,\dots,|j_{max}-j_p|$.
Here $\cE (t,\ta)$ is the real electric field measured in $W^{1/2}$, 
which dependence on the transverse coordinates has been factored out.
Introducing pump laser frequency $\Omega$, we define mode detunings 
\be 
\delta_\mu=\omega_\mu-\Omega,\label{freq2}
\ee 
where $\delta_0$ is the detuning for mode $j_p$, and  
\be
\om_\mu=\om_0+ D_1\mu+\tfrac{1}{2!}D_2\mu^2+\tfrac{1}{3!}D_3\mu^3+\dots\label{freq1},
\ee
where a desired number of the dispersion orders can be included to approximate $\om_\mu$ over a required spectral range. 

$D_1$ is the resonator repetition rate (or free spectral range (FSR)) and $D_2$ is its group velocity dispersion.
$D_2>0$ implies anomalous and $D_2<0$ normal dispersion.
For example, the work  \cite{vahala} deals with
a bi-directionally pumped silica ring with radius $1.5$mm and it has 
$D_1= 2\pi\times 22$GHz, $D_2=2\pi\times 16$kHz. The linewidth of this resonator is $\kappa=2\pi\times 1.5$MHz, and hence the corresponding  finesse $\cF=D_1/\kappa\simeq 13000$. The mode area  estimate is $S_{j_p}\simeq 30\mu$m$^2$, which gives $b_j^2\simeq 4\times 10^{12}$V$^2$W$^{-1}$m$^{-2}$. Pump laser  wavelength was  $\simeq 1550$nm ($\om_0\simeq 2\pi\times 193$THz) and the comb spectra observed there were relatively narrow and span over $\sim 20$nm bandwidth, corresponding to about $300$ modes, and the momentum of a mode nearest to the pump is estimated as $j_p=8700$.

In order to introduce a new set of mode amplitudes important in what follows, we transform Eq. \bref{field1} further:
\begin{subequations}
	\label{field2}
	\begin{align}    
	\cE &=
	e^{ij_p\ta-i\Omega t}\sum_{\mu}B^+_\mu e^{-i\delta_\mu t+i\mu\ta}+
	e^{ij_p\ta+i\Omega t}\sum_{\mu}B^-_\mu e^{i\delta_\mu t+i\mu\ta}+c.c.
	\label{field2a}\\
	&=e^{ij_p\ta-i\Omega t}\sum_{\mu}Q^+_\mu e^{i\mu\ta}+
	e^{ij_p\ta+i\Omega t}\sum_{\mu}Q^{-*}_\mu e^{i\mu\ta}
	+c.c.
	\label{field2b} \\
	&= e^{ij_p\ta-i\Omega t}Q_++
	e^{ij_p\ta+i\Omega t}Q_-^{*}+c.c.~.
	\label{field2c}
	\end{align}
\end{subequations}
Newly introduced mode amplitudes $Q_\mu^\pm$ are  defined as  
\be 
Q_\mu^{+}=B_\mu^{+}e^{- i\delta_\mu t},~Q_\mu^{-}=B_\mu^{-*}e^{-i\delta_\mu t},
\label{qdef}
\ee
and as we can see they absorb frequency scales associated with both $D_1$ and $D_2$. The corresponding CW and CCW envelope functions are
\be 
Q_{\pm}(t,\ta)=\sum_\mu Q_\mu^{\pm}e^{\pm i\mu\ta}.
\label{qdef1}
\ee
Inclusion of the backscattering effects to the envelope equations, see Section 5, requires  introducing of the  envelope functions with  reflections of their spatial coordinate,
\begin{subequations}
	\label{reflect1}
	\begin{align} 
	&
	Q_{\pm}^{(r)}(t,\ta)=\sum_\mu Q_\mu^{\pm}e^{\mp i\mu\ta},\\
	&
	Q^{(r)}_\pm(t,\ta)=Q_\pm(t,2\pi -\ta).
	\end{align}
\end{subequations}
The above definition of the space reflected functions follows a text-book list of  properties of Fourier transforms, where  an equivalent transformation is typically introduced in time domain and could be called as either time reflection or time inversion transformation. Differential equations involving functions with reflections of their arguments also attracted some attention from a more general mathematics prospective, see, e.g., \cite{math}, while our system reveals their role in nonlinear photonics.

In order to take control of $D_1$ in our future calculations, we define yet another set of slow amplitudes 
\begin{subequations}
	\label{def_a}
	\begin{align} 
	&\label{def_aa}
	A_\mu^{+}=
	B_\mu^{+}e^{- i\td t},~  A_\mu^{-}= B_\mu^{-*}e^{- i\td t},\\
	&
	\td=\delta_0+\tfrac{1}{2}D_2\mu^2.
	\label{def_ab} 
	\end{align}
\end{subequations}
Here $D_1$ is moved away from the exponential factors defining our third and final set of amplitudes $A_\mu^\pm$. Instead, exponents with $D_1$ appear explicitly in the total field equation that uses  $A_\mu^\pm$,
\be
\cE = \Big(
e^{ij_p\ta-i\Omega t}\sum_{\mu}A^+_\mu e^{i\mu\big(\ta-D_1t\big)}+
e^{ij_p\ta+i\Omega t}\sum_{\mu}A^{-*}_\mu e^{i\mu\big(\ta+D_1t\big)}\Big)+c.c.~.
\label{field3}
\ee
We also use $A_\mu^\pm$ to define the corresponding envelope functions and their reflections
\begin{subequations}
	\label{envel_a}
	\begin{align} 
	\label{envel_aa} &A_{\pm}=\sum_\mu A_\mu^{\pm}e^{\pm i\mu\ta},~ 
	A_{\pm}^{(r)}=\sum_\mu A_\mu^{\pm}e^{\mp i\mu\ta},\\
	\label{envel_ab} & A^{(r)}_\pm(t,\ta)=A_\pm(t,2\pi -\ta).
	\end{align}
\end{subequations}
Though the envelopes $A_\pm$ can not be used themselves 
to define the electric field $\cE$ (only their mode amplitudes can), cf. Eq. \bref{field3} and \bref{envel_a},  
they play a pivotal role in the transition  from 
the coupled mode to the partial differential equations, see Section 5.

To summarize this section:   $B_\mu^{\pm}$ amplitudes absorb only the slowest time scales associated with the nonlinear effects and resonator losses. $A_\mu^{\pm}$  absorb  time scales associated with the second and higher order dispersions, in addition to the ones already inside $B_\mu^{\pm}$. $Q_\mu^{\pm}$ amplitudes evolve with the highest in our hierarchy frequency determined by the resonator repetition rate. To see how these different time scales and mode amplitudes are used to express the  total field, $\cE$, one should compare Eqs. \bref{field2a}, \bref{field2b} and \bref{field3}.   

The rest of this work is structured as follows: In  section 3, we first derive a system of equations for $B^{\pm}_\mu$ and perform its exact reduction to the equations for $Q^{\pm}_\mu$. In Section 4, we
come back to the equations for $B^{\pm}_\mu$, make the $D_1$ role explicit, 
eliminate the associated fast oscillations and derive a simpler system for $A^{\pm}_\mu$. Corresponding mean-field  equations for the envelope functions $Q_{\pm}$ and $A_{\pm}$ and their counter parts with the reflected spatial coordinates are  derived in Section 5.

\section{Coupled mode equations}

\subsection{Separating equations for CW and CCW amplitudes}
Substituting  $t'=t-\tau$ in Eqs. \bref{max}, \bref{field} we then assume that material response is fast so that $C_j(t-\tau)\simeq C_j(t)-\tau\p_t C_j+\dots$. Neglecting all the 2nd and higher order time derivatives of  $B^\pm_\mu$ we find that Eq. (\ref{max}) transforms to
\begin{align}
	\nn
	&-\p_t^2\cN \simeq
	b_{j_p}F_{ j_p}\sum_\mu e^{ij_\mu \ta}\times\Big(-
	n_\mu^2\om_\mu^2\ep_{in}(\ta)B^+_\mu e^{-i\om_\mu t}
	-2i\om_\mu s_\mu e^{-i\om_\mu t}\p_tB^+_\mu
	\nn \\&-n_\mu^2\om_\mu^2\ep_{in}(\ta)B^-_\mu e^{i\om_\mu t}+2i\om_\mu s_\mu e^{i\om_\mu t}\p_tB^-_\mu \Big)\nn \\
	&
	+b_{j_p}F_{j_p}\sum_\mu e^{-ij_\mu \ta}\times\Big(
	-  n_\mu^2\om_\mu^2\ep_{in}(\ta)B^{+*}_\mu e^{i\om_\mu t}+2i\om_\mu s_\mu e^{i\om_\mu t}\p_tB^{+*}_\mu
	\nn\\ &-n_\mu^2\om_\mu^2\ep_{in}(\ta)B^{-*}_\mu e^{-i\om_\mu t}-2i\om_\mu s_\mu e^{-i\om_\mu t}\p_tB^{-*}_\mu\Big),
	\label{left}
	\end{align}
where $s_\mu=n_\mu^2+\frac{1}{2}\om_\mu\p_\om n_\mu^2\simeq n_\mu^2$. We then expand nonlinear polarization $\cN$ in Fourier series
\be 
\cN(r,z,\ta,t)=\sum_\mu\cN_{ j_\mu}(r,z,t)e^{ij_\mu\ta}+c.c. .
\label{na}
\ee
In order to carry out separation of the CW and CCW equation we also need to define CW and CCW components of nonlinear polarization, $P^\pm_{j_\mu}e^{\mp i\om_{j_\mu} t}$, such that
\be 
\cN_{j_\mu}\equiv P^+_{j_\mu}e^{-i\om_{j_\mu} t}+P^-_{ j_\mu}e^{i\om_{j_\mu} t}.
\label{nc}
\ee
Explicit expressions for $P^\pm_{j_\mu}$ are given by Eqs. \bref{nd} below.

We  now multiply the left and right hand-sides of Eq. (\ref{left}) by
$b_{j_p}F_{j_p}\exp^{-ij_{\mu'}\ta}$, integrate in $r,z$ and $\ta$, and
approximate  $\om_\mu\simeq\om_0=\om_{j_p}$, $n_\mu\simeq n_0$ inside all the pre-factors, but not in the powers of the exponents.
The resulting model, see Eqs. (\ref{mod1a}), makes use of the two  scattering matrices having dimensions of angular frequencies. 
One characterises scattering induced coupling between the co-propagating modes
\be\wh\Gamma_{\mu\mu'}=\tfrac{1}{2}\om_0
\int_0^{2\pi}e^{i(\mu-\mu')\ta}
\ep_{in}(\ta) \frac{d\ta}{2\pi} ,\ee
and the other one describes backscattering induced mode coupling,
\be\wh\cR_{\mu\mu'}=\tfrac{1}{2}\om_0
\int_0^{2\pi} e^{-i(2j_p+\mu+\mu')\ta}\ep_{in}(\ta) \frac{d\ta}{2\pi}.
\ee

The projected equation itself is 
	\ba \nn
	&&-\sum_\mu\wh\Gamma_{\mu\mu'}B^+_\mu e^{-i\om_\mu t}-ie^{-i\om_{\mu'} t} \p_tB^+_{\mu'}
	-\sum_\mu\wh\Gamma_{\mu\mu'}B^-_\mu e^{i\om_\mu t}+ie^{i\om_{\mu'} t} \p_tB^-_{\mu'}
	\nn \\ \nn &&-\sum_\mu\wh\cR_{\mu\mu'}B^{+*}_\mu e^{i\om_\mu t} 
	-\sum_\mu\wh\cR_{\mu\mu'}B^{-*}_\mu e^{-i\om_\mu t} \\&& 
	=-\frac{\pi\om_0}{n_0^2V_pb_{j_p}^2}\p_t^2 \iint \cN_{j_{\mu'}}b_{j_p}F_{j_p}rdrdz\nn \\ &&\simeq
	\frac{\pi\om_0}{n_0^2V_pb_{j_p}^2}\iint\Big( P^+_{j_\mu'}e^{-i\om_{\mu'} t}+P^-_{j_\mu'}e^{i\om_{\mu'} t}\Big)b_{j_p}F_{j_p}rdrdz.\label{mod1a}
	\ea
where    
$V_p=2\pi \iint~  F_{j_p}^2 rdrdz$ is the mode volume for $j=j_p$. 

Eq. \bref{mod1a}  can now be split, as per rotating wave approximation, 
into the parts proportional to $e^{\pm i\om_{j_\mu}t}$ exponents, so that we have two equations defined on the slow, $D_2$ related, time scales:
\begin{subequations}
	\label{mod2}
	\begin{align}
		-i \p_tB^+_{\mu}
	=\sum_{\mu'}\Big(\wh\Gamma_{\mu'\mu}B^+_{\mu'}+\wh\cR_{\mu'\mu}B^{-*}_{\mu'}\Big)e^{i(\om_{\mu} -\om_{\mu'})t}
	+\frac{\pi \om_0}{n_0^2V_pb_{j_p}^2}\iint P^+_{j_\mu}b_{j_p}F_{j_p}rdrdz, &\\
		i \p_tB^{-}_{\mu}
	=\sum_{\mu'}\Big(\wh\Gamma_{\mu'\mu}B^{-}_{\mu'}+\wh\cR_{\mu'\mu}B^{+*}_{\mu'}\Big)e^{-i(\om_{\mu}-\om_{\mu'}) t} +\frac{\pi \om_0}{n_0^2V_pb_{j_p}^2}\iint P^{-}_{j_\mu}b_{j_p}F_{j_p}rdrdz,&
	\end{align}
\end{subequations}
where we have also swapped $\mu$ and $\mu'$. 

In order to be used to describe laboratory experiments with microresonators, Eqs. \bref{mod2} have to be amended with the single mode pump term  and losses accounting for the finite linewidth. 
We take, for the laser frequency at the exact cavity resonance $\Omega=\om_{\mu=0}=\om_{j=j_p}$ and for the low pump levels, i.e., linear regime, the intracavity powers of CW and CCW waves to be $|\cH_\pm|^2=|B^\pm_\mu|^2$. This  is achieved via a phenomenological substitution 
\be
\label{pump1}
i\p_tB^{\pm}_\mu\to  i\p_t B^\pm_\mu + i\tfrac{1}{2}\kappa
\left( B^\pm_\mu-\wh\delta_{\mu, 0}\cH_\pm e^{\pm i(\omega_\mu-\Omega)t}\right).
\ee
Here,  Kronecker delta is defined as $\wh\delta_{\mu, \mu_1}=1$ for $\mu=\mu_1$ and is $0$ otherwise.  

If pump is absent, then the field power would decay with the rate $\kappa$ (full width of the resonance). An expression linking $\cH_\pm$ with the laser powers $\cW_\pm$ is
\begin{equation}
|\cH_\pm|^2=\frac{\eta}{\pi}\cF  \cW_\pm, 
\end{equation}
where $\cW_\pm$ are the laser powers pumping, respectively, CW and CCW waves. $\eta<1$ is the coupling efficiency via, e.g., a prism or a waveguide, into a resonator mode. $\eta=\kappa_{c}/\kappa$, where $\kappa_{c}$ is the coupling pump rate (equals coupling loss rate). $\cF/\pi$ is the cavity induced power enhancement. Detailed theoretical and experimental studies of the power enhancement effect and coupling in and out considerations for ring cavities can be found in, e.g.,  \cite{book1,book2}. 

$\wh\cR_{\mu\mu}\sim 2\pi\times 4$~kHz  in  Ref. \cite{vahala}. In this regime, it is safe to assume
that $\kappa$ dominates over $\wh\Gamma$ and $\wh\cR$ terms. 
Using this we disregard  $\wh\Gamma_{\mu'\mu}$ in what follows, and retain only the dominant diagonal terms in
$\wh\cR_{\mu'\mu}$, i.e., $\wh\cR_{\mu'\mu\ne\mu'}\approx0 $. Dispersion of the diagonal terms  is also disregarded, 
$\wh\cR_{\mu\mu}\simeq\wh\cR_{00}= R$. 
Accounting for all of the above and complex conjugating second of  Eqs. \bref{mod2}  we conclude this subsection with
\begin{subequations}
	\label{mod3}
	\begin{align} 
	&
	i\p_t B^+_{\mu}= -i\tfrac{1}{2}\kappa\left( B^+_\mu
	-\wh\delta_{\mu, 0}\cH _+e^{i\delta_\mu t}\right)
	-R B^{-*}_\mu-\frac{\pi\om_0 }{n_0^2V_pb_{j_p}^2}\iint P_{j_\mu}^+b_{j_p}F_{j_p}rdrdz,\\ 
	& i \p_t B^{-*}_\mu=-i\tfrac{1}{2}\kappa\left( B^{-*}_\mu
	-\wh\delta_{\mu, 0}\cH _-e^{i\delta_\mu t}\right)
	-R^* B^{+}_\mu -\frac{\pi\om_0 }{n_0^2V_pb_{j_p}^2}\iint P_{j_\mu}^{-*}b_{j_p}F_{j_p}rdrdz.
	\end{align}
\end{subequations}

\subsection{Opening up nonlinearity}
Using Eqs. \bref{n3}, \bref{field1} we have 
\begin{equation}
\cN= b_{j_p}^3F_{j_p}^3\chi^{(3)}\cE^3,
\label{n33}
\end{equation} 
and
\be 
\cE^3= 3\Big\{e^{ij_p\ta-i\Omega t}\left(|Q_+|^2+2|Q_-|^2\right)Q_+ 
+e^{-ij_p\ta-i\Omega t}\left(|Q_-|^2+2|Q_+|^2\right)Q_- +\dots\Big\}+c.c.~.
\label{n1}
\ee
Comparing Eqs. \bref{n33}, \bref{n1} and Eqs. \bref{na}, \bref{nc}, one can define explicit expressions for $P^\pm_{j_\mu}$.
Assuming spectrally narrow combs, and therefore omitting all terms with exponential factors 
oscillating in space with multiples of $j_p$ and in time with multiples of $\Omega$, we find
\begin{subequations}
	\label{nd}
	\begin{align} 
	&  P_{j_\mu}^+=3b_{j_p}^3F_{j_p}^3\chi^{(3)}e^{i(\om_{j_\mu}-\Omega)t}
	\int_0^{2\pi}\left(|Q_+|^2+2|Q_-|^2\right)Q_+e^{-i\mu\ta}
	\frac{d\ta}{2\pi},\\
	&  P_{j_\mu}^-=3b_{j_p}^3F_{j_p}^3\chi^{(3)}e^{-i(\om_{j_\mu}-\Omega)t}
		\int_0^{2\pi}\left(|Q_-|^2+2|Q_+|^2\right)Q_-^*e^{-i\mu\ta}
	\frac{d\ta}{2\pi}.
	\end{align}
\end{subequations}
Thereby, Eqs. \bref{mod3}  become
\begin{subequations}
	\label{mod4}
	\begin{align} &
	 i\p_t B^+_{\mu}= -i\tfrac{1}{2}\kappa\left( B^+_\mu
	-\wh\delta_{\mu, 0}\cH _+e^{i\delta_\mu t}\right)
	-R B^{-*}_\mu -\gamma e^{i\delta_{\mu}t}
	\int_0^{2\pi}\left(|Q_+|^2+2|Q_-|^2\right)Q_+e^{-i\mu\ta}
	\frac{d\ta}{2\pi},\label{mod4a}\\ 
	& i \p_t B^{-*}_\mu=-i\tfrac{1}{2}\kappa\left( B^{-*}_\mu
	-\wh\delta_{\mu, 0}\cH _-e^{i\delta_\mu t}\right)
	-R^* B^{+}_\mu -\gamma e^{i\delta_\mu t}
	\int_0^{2\pi}\left(|Q_-|^2+2|Q_+|^2\right)Q_-e^{i\mu\ta}
	\frac{d\ta}{2\pi},\label{mod4b} 
	\end{align}
\end{subequations}
where nonlinear coefficient is
\be\label{gamma}
\gamma=\frac{3}{2}\frac{\om_0 b_{j_p}^2}{n_0^2}\frac{2\pi}{V_p}\iint \chi^{(3)}F_{j_p}^4rdrdz.
\ee
Total refractive index, $n$, for a single mode operation is $n=\{n_{0}^2+3\chi^{(3)}b_{j_p}^2|B^{+}_{j_p}|^2\}^{1/2}\simeq n_{0}+\tfrac{3}{2n_0}\chi^{(3)}b_{j_p}^2|B^{+}_{j_p}|^2=n_0+n_2I^{+}_{j_p}$, see definition of intensity  before Eq. \bref{b}. Hence, Kerr coefficient is $n_2=\tfrac{3}{4}\chi^{(3)}(n_0^2\epsilon_{vac} c)^{-1}$. Using Eqs. \bref{b}, \bref{gamma}, an  expression for $\gamma$ in terms of more often used $n_2$ is
\be\label{gamma1}
\gamma= \frac{\om_0}{S_{j_p}n_0}\frac{2\pi}{V_p}\iint n_2F_{j_p}^4rdrdz.
\ee
Assuming that the $j_p$ mode is well confined within the resonator material, the mode 
shape can be approximated by a Gaussian function (allowing for different widths along $z$ and $x$), and $rdr\approx r_0dx$ (see text before Eq. \bref{b}),  gives $2\pi\iint F_{j_p}^4 rdrdz/V_p\approx\tfrac{1}{2}$ and
\be\label{gamma2}
\gamma\approx \frac{\om_0n_2}{2S_{j_p}n_0}.
\ee
Eq. \bref{gamma1} and Eq. \bref{gamma2} have been compared using mode profiles calculated with Comsol and it was found that the latter provides a very practical approximation.   
For $\om_0=2\pi\times 193$THz, $n_0=1.47$ and $n_2\simeq 3.2\times 10^{-20}$m$^2$/W$^2$ (silica glass), and mode area $S_{j_p}\approx 30\mu$m$^2$  we have $\gamma\simeq 2\pi\times 70$kHz/W. $S_{j_p}$ is an order of magnitude smaller and $n_2$ is an order of magnitude larger in integrated Si$_3$N$_4$ microresonators, and their combined effect boosts $\gamma$ up by two orders of magnitude.

Using Eqs. \bref{qdef} to express amplitudes $B_\mu^\pm$ via $Q_\mu^{\pm}$ we find that all the time dependent exponents cancel out and the resulting coupled mode equations for $Q_\mu^\pm$ amplitudes are 
\begin{subequations}
	\label{mod5a}
	\begin{align} 
	&
	 i\p_t Q^+_{\mu}=\delta_\mu Q^+_{\mu} -i\tfrac{1}{2}\kappa\left( Q^+_\mu
	-\wh\delta_{\mu, 0}\cH_+\right)
	-R Q^{-}_\mu-\gamma 
	\int_0^{2\pi}\left(|Q_+|^2+2|Q_-|^2\right)Q_+e^{-i\mu\ta}
	\frac{d\ta}{2\pi},\label{mod5aa}\\
	& i \p_t Q^{-}_\mu=\delta_\mu Q^-_{\mu} -i\tfrac{1}{2}\kappa\left( Q^{-}_\mu
	-\wh\delta_{\mu, 0}\cH_-\right)
	-R^* Q^{+}_\mu-\gamma 
	\int_0^{2\pi}\left(|Q_-|^2+2|Q_+|^2\right)Q_-e^{i\mu\ta}
	\frac{d\ta}{2\pi}, \label{mod5ab}
	\end{align}
\end{subequations}
where $Q_\pm$ envelopes are given by Eqs. \bref{qdef1}.

\section{Washout of the repetition rate timescales from the coupled mode equations}
Systems of Eqs. \bref{mod4}, \bref{qdef}, \bref{qdef1} on one side, and Eqs. \bref{mod5a}, \bref{qdef1} on the other, are mathematically and physically equivalent. However, there are important observations to be made here.
If one could assume that $|Q_+|^2+2|Q_-|^2$ under the integrals in the right hand sides of Eqs. \bref{mod4} and Eqs. \bref{mod5a} is a slow function of time, then these integrals would be approximately equal to $Q_{\mu}^{\pm}e^{-i\delta_\mu t}$, see Eqs. \bref{qdef}. Balancing these with the $e^{i\delta_\mu t}$ exponents before the integrals in Eqs. \bref{mod4},  one would end up with equations 
involving  time scales determined only by the linewidths, pump detuning and nonlinear resonance shifts, which are all order of MHz. MHz frequencies would be far simpler to resolve numerically, compare to GHz-THz frequencies associated with $D_1$, that are directly implicated inside $\delta_\mu$ in the  linear parts of Eqs. \bref{mod5a}. 

In this Section, we demonstrate that there are both slow and fast time scales 
inside the nonlinear terms in Eqs. \bref{mod4}, and that the latter can be eliminated resulting in a simpler and better balanced system of equations for the $A_\mu^{\pm}$ amplitudes, see Eqs. \bref{def_a}, \bref{mod6d}.

We proceed by taking  Eqs. \bref{mod4a}, express $Q_{\pm}$ via  $B_\mu^{\pm}$, see Eqs. \bref{qdef}, \bref{qdef1}, calculate integrals in the nonlinear terms, see Eq. \bref{long_a}, and perform the two step transformation, see Eqs. \bref{long_b}, \bref{long_c},
	\begin{subequations}
		\label{bb}
		\begin{eqnarray} 	
		&& i\p_t B^+_{\mu}+i\tfrac{1}{2}\kappa\left(B^+_\mu
		-\wh\delta_{\mu, 0}\cH_\pm e^{i\delta_\mu t}\right)+R B^{-*}_\mu= \nn \\ &&
				-\gamma e^{i\delta_\mu t}\sum_{\mu_1\mu_2\mu_3}\Big(
		\wh\delta_{\mu_1+\mu_2-\mu_3,\mu}
		B_{\mu_1}^+B_{\mu_2}^+B_{\mu_3}^{+*}
		e^{i(-\delta_{\mu_1}-\delta_{\mu_2}+\delta_{\mu_3})t}\nn\\ && 
		+2\wh\delta_{\mu_1-\mu_2+\mu_3,\mu}
		B_{\mu_1}^+B_{\mu_2}^{-}B_{\mu_3}^{-*}
		e^{i(-\delta_{\mu_1}-\delta_{\mu_2}+\delta_{\mu_3})t}
		\Big)=
		\label{long_a} \\&&
				-\gamma \sum_{\mu_1\mu_2\mu_3}
		\wh\delta_{\mu_1+\mu_2-\mu_3,\mu}\Big(
		B_{\mu_1}^+B_{\mu_2}^+B_{\mu_3}^{+*}
		e^{i(-\delta_{\mu_1}-\delta_{\mu_2}+\delta_{\mu_3}+\delta_\mu)t}\nn \\ &&
		+2B_{\mu_1}^+B_{\mu_3}^{-}B_{\mu_2}^{-*}
		e^{i(-\delta_{\mu_1}-\delta_{\mu_3}+\delta_{\mu_2}+\delta_\mu)t}
		\Big)=
		\label{long_b} \\ &&
				-\gamma \sum_{\mu_1\mu_2\mu_3}
		\wh\delta_{\mu_1+\mu_2-\mu_3,\mu}\Big(
		B_{\mu_1}^+B_{\mu_2}^+B_{\mu_3}^{+*}
		e^{\tfrac{iD_2}{2}(\mu^2-\mu_1^2-\mu_2^2+\mu_3^2)t}\nn \\&& 
		+2B_{\mu_1}^+B_{\mu_2}^{-*}B_{\mu_3}^{-}
		e^{i2D_1(\mu_2-\mu_3)t}e^{\tfrac{iD_2}{2}(\mu^2-\mu_1^2-\mu_3^2+\mu_2^2)t}
		\Big).
		\label{long_c}
		\end{eqnarray}
	\end{subequations} 
The four-wave mixing momentum matching conditions are reflected in the Kronecker delta's in front of the nonlinear terms in the second line of the above and directly follow from  taking the integrals in $\theta$.
Swapping of $\mu_2$ and $\mu_3$ inside the nonlinear cross-coupling is a critical step that a reader should pay attention to, see Eq. \bref{long_b}. This operation equals the Kronecker delta's, but it re-orders the amplitudes and respective frequency detunings in the second nonlinear term.
After inserting explicit expressions for $\delta_\mu$, see Eqs. \bref{freq2}, \bref{freq1}, and using  
the momentum matching condition, \be\mu_1+\mu_2=\mu_3+\mu,\ee we find that $D_1$ frequencies cancel out inside the nonlinear self-action terms, but remain in the cross-action ones providing $\mu_2\ne\mu_3$, see Eq. \bref{long_c}. Thus if $D_1$ oscillations are much faster than dynamics associated with the other time scales left in the equations, i.e., $\kappa$, $R$ and nonlinear frequency shifts, then the fast oscillating components can be disregarded  \cite{lobanov,cole}. This leaves us only with $\mu_2=\mu_3$ components in the cross-action terms, so  that
\begin{align}
& 
i\p_t B^+_{\mu}+i\tfrac{1}{2}\kappa\left(B^+_\mu
-\wh\delta_{\mu, 0}\cH_\pm e^{i\delta_\mu t}\right)+R B^{-*}_\mu= \nn \\
&-\gamma\sum_{\mu_1\mu_2\mu_3}
\wh\delta_{\mu_1+\mu_2-\mu_3,\mu}
B_{\mu_1}^+B_{\mu_2}^+B_{\mu_3}^{+*}
e^{\tfrac{iD_2}{2}(\mu^2-\mu_1^2-\mu_2^2+\mu_3^2)t} 
-2\gamma B_{\mu}^+\sum_{\mu_2}|B_{\mu_2}^{-}|^2.\label{mod6b}
\end{align}
Nonlinear terms in  Eqs. \bref{mod6b} are now grouped into the phase insensitive pure cross-Kerr term, that contains  nonlinear shift of the CW resonance frequencies due to CCW wave, and into the term that mixes both phase sensitive and phase insensitive four-wave mixing CW-CW nonlinearities. The phase sensitive effects come only from the CW-CW interaction, because all the phase sensitive CW-CCW  dynamics develops with the $2D_1$ frequencies and is washed out by the high repetition rates. This can be called the {\em washout effect} of high repetition rates on nonlinear frequency mixing of the counter-propagating waves in a ring resonator. 

Using   $A_{\mu}^{\pm}$ amplitudes and detunings $\delta'_\mu$, which are both $D_1$ free, see  Eqs. \bref{def_a}, allows to hide $e^{iD_2\mu^2t/2}$ exponents in Eqs. \bref{mod6b}. Adding the CCW equation, we have
\begin{subequations}
	\label{mod6c}
	\begin{align}
	& 
	i\p_t A^+_{\mu}-\delta'_\mu A^+_{\mu}+i\tfrac{1}{2}\kappa\left(A^+_\mu
	-\wh\delta_{\mu, 0}\cH_\pm \right)+R A^{-}_\mu= \nn \\
	&
	-\gamma \sum_{\mu_1\mu_2\mu_3}
	\wh\delta_{\mu_1+\mu_2-\mu_3,\mu}
	A_{\mu_1}^+A_{\mu_2}^+A_{\mu_3}^{+*}
	-2\gamma A_{\mu}^+\sum_{\mu_2}|A_{\mu_2}^{-}|^2,\\
	& 
	i\p_t A^-_{\mu}-\delta'_\mu A^-_{\mu}+i\tfrac{1}{2}\kappa\left(A^-_\mu
	-\wh\delta_{\mu, 0}\cH_\pm \right)+R^* A^{+}_\mu= \nn \\
	&
	-\gamma \sum_{\mu_1\mu_2\mu_3}
	\wh\delta_{\mu_1+\mu_2-\mu_3,\mu}
	A_{\mu_1}^-A_{\mu_2}^-A_{\mu_3}^{-*}
	-2\gamma A_{\mu}^-\sum_{\mu_2}|A_{\mu_2}^{+}|^2.
	\end{align}
\end{subequations}
The difference of the above nonlinear terms with the ones in the equations for $Q^\pm_\mu$, see Eq. \bref{mod5a}, that include un-averaged $D_1$ oscillations, becomes more obvious, if the sums in Eqs. 
\bref{mod6c} are replaced with the integrals, see also Eq. \bref{envel_aa},
\begin{subequations}
	\label{mod6d}
	\begin{align}
	& 
	i\p_t A^+_{\mu}-\delta'_\mu A^+_{\mu}+i\tfrac{1}{2}\kappa\left(A^+_\mu
	-\wh\delta_{\mu, 0}\cH_\pm \right)+R A^{-}_\mu= \nn \\
	&
	-\gamma \int_0^{2\pi} |A_+|^2A_+ e^{-i\mu\ta} \frac{d\ta}{2\pi}
	-2\gamma A_\mu^+ \int_0^{2\pi}
	|A_-|^2 \frac{d\ta}{2\pi},\\ \nn
	& 
	i\p_t A^-_{\mu}-\delta'_\mu A^-_{\mu} + i\tfrac{1}{2} \kappa \left(A^-_\mu
	-\wh\delta_{\mu, 0}\cH_\pm \right)+R^* A^{+}_\mu= \nn \\
	&-\gamma \int_0^{2\pi} |A_-|^2A_- e^{i\mu\ta} \frac{d\ta}{2\pi}
	-2\gamma A_\mu^- \int_0^{2\pi}
	|A_+|^2 \frac{d\ta}{2\pi}.
	\end{align}
\end{subequations}
The last terms in Eqs. \bref{mod6d} follow from the Parseval's theorem.  Thus Eqs. \bref{mod6b}
include only effects of the second and higher order dispersions in both linear and nonlinear terms, that in microresonators are associated with the kHz to MHz time scales. Hence solving 
Eqs. \bref{mod6b} is expected to provide significant computational advantages over all other versions of the coupled mode equations. 

\section{Envelope models}
Connection of the coupled mode equations to the wave dynamics becomes more intuitive, if one now derives the envelope,
Lugiato-Lefever like, equations. First, we take the $Q_\mu^{\pm}$ model,  see Eqs. \bref{mod5a}, and
multiply Eq. \bref{mod5aa} with $e^{i\mu\ta}$ and \bref{mod5ab} with $e^{-i\mu\ta}$.
We then sum up each of the equations in $\mu$ and use Eqs. \bref{qdef1}, \bref{reflect1} connecting the envelopes $Q_{\pm}$ and the reflected envelopes $Q_{\pm}^{(r)}$ to their mode amplitudes. This procedure is free from approximations and it leads to a system of partial differential equations for  $Q_\pm$ and $Q_\pm^{(r)}$,
\begin{subequations}
	\label{ll1}
	\begin{align}
	& i\p_tQ_+=\delta_0 Q_++\left(-iD_1\p_\ta -\tfrac{1}{2!}D_2\p_\ta^2+i\tfrac{1}{3!}D_3\p_\ta^3+\dots\right) Q_+-RQ_-^{(r)}\nn \\
	&-i\tfrac{1}{2}\kappa (Q_+-\cH_+)-\gamma (|Q_+|^2+2|Q_-|^2)Q_+,\label{ll1a}\\
	& i\p_tQ_-=\delta_0 Q_-+\left(+iD_1\p_\ta -\tfrac{1}{2!}D_2\p_\ta^2-i\tfrac{1}{3!}D_3\p_\ta^3+\dots\right) Q_--R^*Q_+^{(r)}\nn \\
	&-i\tfrac{1}{2}\kappa (Q_--\cH_-)-\gamma (|Q_-|^2+2|Q_+|^2)Q_-. \label{ll1b}
	\end{align}
\end{subequations}
To form a closed system, the above pair of equations should be supplemented with two more equations for the $Q_\pm^{(r)}$, see Eqs. \bref{reflect1} defining 
$\ta$ reflection. 

Starting from the equations for  $A_\mu^{\pm}$, see Eqs. \bref{mod6d},  we follow a modified procedure. Namely, we multiply both CW and CCW equations by the same exponent $e^{i\mu\ta}$,  use the envelope definitions in Eqs. \bref{envel_a},
observe that $\int_0^{2\pi}|A_-|^2A_-e^{i\mu\ta}d\ta=\int^0_{-2\pi}|A_-^{(r)}|^2A_-^{(r)}e^{-i\mu\ta}d\ta$ and, due to periodicity, $=\int_0^{2\pi}|A_-^{(r)}|^2A_-^{(r)}e^{-i\mu\ta}d\ta$, sum up in $\mu$, and derive the following envelope equations
\begin{subequations}
	\label{ll2}
	\begin{align}
	& i\p_tA_+=\delta_0A_++\left(-\tfrac{1}{2!}D_2\p_\ta^2+i\tfrac{1}{3!}D_3\p_\ta^3 +\dots\right)A_+-RA_-^{(r)}-\gamma |A_+|^2A_+\nn \\
	&-i\tfrac{1}{2}\kappa (A_+-\cH_+)-2\gamma A_+\int_0^{2\pi}
	|A_-^{(r)}(\ta')|^2 \frac{d\ta'}{2\pi},\label{ll2a}\\
	& i\p_tA_-^{(r)}=\delta_0A_-^{(r)}+\left(-\tfrac{1}{2!}D_2\p_\ta^2+i\tfrac{1}{3!}D_3\p_\ta^3 +\dots\right) A_-^{(r)}-R^*A_+-\gamma |A_-^{(r)}|^2A_-^{(r)}\nn \\
	&-i\tfrac{1}{2}\kappa (A_-^{(r)}-\cH_-)-2\gamma A_-^{(r)}\int_0^{2\pi}
	|A_+(\ta')|^2 \frac{d\ta'}{2\pi}. \label{ll2b}
	\end{align}
\end{subequations}
The above equations do not only exclude the $D_1$ dynamics, but also form a closed system of two equations for the CW $A_+$
envelope and for the reflected  CCW $A_-^{(r)}$ envelope, see Eqs. \bref{envel_a}. They can also be supplemented
with equations for $A_+^{(r)}$, $A_-$, but this time those are left as an independent pair. Again numerical modelling of Eqs. \bref{ll2} is expected to have great advantages relative to working with Eqs. \bref{ll1}.  
Similar to ours procedure to remove the $D_1$ linked time scales has been developed  for the Kerr Fabry-Perot cavities supporting a single family of standing waves and hence yielding a one-component Lugiato-Lefever model \cite{cole}. The respective ring geometry model in \cite{lobanov}  mixes all four envelope functions, i.e., $A_\pm$, $A_\pm^{(r)}$, and is limited by the second order dispersion. 

Eqs. \bref{ll1} (not Eqs. \bref{ll2}) could in fact, be written without a rigorous derivation, by simply  relying on common knowledge, let aside reflected envelopes in the backscattering terms. These equations include traditional cross-phase modulation, and also repetition rates terms and other odd order dispersion terms with the opposite signs. Contrary, Eqs. \bref{ll2} have no repetition rate terms, i.e., $D_1$-terms, and the remaining odd dispersions, i.e., $D_3$, $D_5$, etc.,  come with the same signs. Simultaneously, phase sensitive nonlinear wave mixing effects induced by CW-CCW  interaction have been washed out. 
The only nonlinear cross-interaction left comes from the integrated power, which merely shifts the detuning parameters.
Thus, in the absence of backscattering a nonlinear bi-directional resonator operates  as a  uni-directional one, but with the detuning parameter altered by the total power of the counter-propagating wave.

\section{Summary}
We have  derived coupled mode equations describing nonlinear wave mixing processes
in Kerr microresonator with counter-propagating waves. 
Features of the first two coupled mode formulations given by  Eqs. \bref{mod4} and Eqs. \bref{mod5a} are that they fully account for the repetition rate effects and that nonlinear terms are taken in the
real space, and can be evaluated via Fourier transforms, see also \cite{wabnitz}.  We then proceeded to present simplified multi-mode equations that neglect the repetition rate dynamics driving the phase sensitive terms responsible for nonlinear interaction between the counter-propagating fields ({\em washout effect}, section 4), and again deal 
with the nonlinearity in the real space, see  Eqs. \bref{mod6d}. 

Finally, we demonstrated that coupled mode equations \bref{mod5a} and Eqs. \bref{mod6d} are equivalent to  two different, Lugiato-Lefever-like, envelope models. The one that involves the repetition rate dynamics, see Eqs. \bref{ll1}, links 
 two usual envelopes for the CW and CCW fields, with two of their space reflections. While the one with the repetition rate averaged out, see Eqs. \bref{ll2}, makes a closed system already for two envelopes, $A_\pm$,  one of which is reflected. We note, that  $Q_\pm$ can be used directly to reconstruct total electric field, see Eq. \bref{field2c}, while $A_\pm$ can not, but their respective mode amplitudes can, see Eqs. \bref{field3}, \bref{envel_aa}. 
 
 We have taken care to reveal all mathematical transformations, that allow a reader to verify our derivation steps and apply modifications if required. Opportunities for future theoretical and numerical studies offered by the models presented here are numerous, as well as their potential to guide and interpret experimental work.

\section*{Funding}
EU Horizon 2020 Framework Programme (812818, MICROCOMB);
UK EPSRC (EP/R008159); Russian
Science Foundation (17-12-01413).

\section*{Acknowledgement}
Discussions with authors of \cite{lobanov} 
and with greatly missed M.L. Gorodetsky are gratefully acknowledged.

\section*{Disclosures}
The author declares no conflicts of interest.

%%%%%%%%%%%%%%%%%%%%%%% References %%%%%%%%%%%%%%%%%%%%%%%%%

\end{document}